\documentclass[namedreferences]{SolarPhysics}
\usepackage[optionalrh]{spr-sola-addons}
\usepackage{solaheader} 
\usepackage{graphicx}        
\usepackage{booktabs}
\usepackage{color}           
\usepackage{url}             
\makeatletter

\newcommand{\Rmnum}[1]{\expandafter\@slowromancap\romannumeral #1@}
\makeatother

\begin{document}

\begin{article}

\begin{opening}

\title{Streamer Wave Events Observed in Solar Cycle 23}

\author{S.W.~\surname{Feng}$^{1,2,3}$\sep
        Y.~\surname{Chen}$^{3}$\sep
        B.~\surname{Li}$^{3}$\sep
        H.Q.~\surname{Song}$^{3}$\sep
        X.L.~\surname{Kong}$^{3}$\sep
        L.D.~\surname{Xia}$^{3}$\sep
        X.S.~\surname{Feng}$^{1}$
       }
\runningauthor{S.W. Feng, \textit{et al.}}
\runningtitle{Streamer Waves}

 \institute{$^{1}$ SIGMA Weather Group, State Key laboratory for Space Weather, Center for Space Science and Applied Research, Chinese Academy of Sciences, Beijing 100190, China \\
                     email: \url{yaochen@sdu.edu.cn}\\
             $^{2}$ College of Earth Sciences, Graduate School of Chinese Academy of Sciences, Beijing 100049, China \\
             $^{3}$ Shandong Provincial Key Laboratory of Optical Astronomy and Solar-Terrestrial Environment,
             School of Space Science and Physics, Shandong University
             at Weihai, Weihai 264209, China
             }

\begin{abstract}
In this paper we conduct a data survey searching for well-defined
streamer wave events observed by the Large Angle and Spectrometric
Coronagraph (LASCO) on-board the Solar and Heliospheric Observatory (SOHO)
throughout Solar Cycle 23. As a result, 8 candidate events are found and
presented here. We compare different events and
find that in most of them the driving CMEs ejecta are characterized by a
high speed and a wide angular span, and the CME-streamer
interactions occur generally along the flank of the streamer
structure at an altitude no higher than the bottom of the field of
view of LASCO C2. In addition, all front-side CMEs have
accompanying flares. These common observational features shed
light on the excitation conditions of streamer wave events.

We also conduct a further analysis on one specific streamer wave
event on 5 June 2003. The heliocentric distances of 4 wave
troughs/crests at various exposure times are determined; they are
then used to deduce the wave properties like period, wavelength,
and phase speeds. It is found that both the period and wavelength
increase gradually with the wave propagation along
the streamer plasma sheet, and the phase speed of the preceding wave
is generally faster than that of the trailing ones. The
associated coronal seismological study yields the radial profiles
of the Alfv\'en speed and magnetic field strength in the region
surrounding the streamer plasma sheet. Both quantities show a
general declining trend with time. This is interpreted as an
observational manifestation of the recovering process of the
CME-disturbed corona. It is also found that the Alfv\'enic
critical point is at about 10 R$_\odot$ where the flow
speed, which equals the Alfv\'en speed, is $\sim$ 200 km s$^{-1}$.
\end{abstract}
\keywords{Coronal Seismology; Waves, Propagation; Magnetic fields, Corona;
Coronal Mass Ejections
 }
\end{opening}

\section{Introduction}
     \label{S-Introduction}
Streamer waves can be excited by the interaction of a rapidly moving and
expanding CME ejecta with a nearby streamer structure, representing
one of the largest wave phenomena ever observed in the solar
corona (Chen \textit{et al}., 2010, Paper \Rmnum{1}). We interpret the waves
as the fast kink body mode (Edwin and Roberts, 1982)
carried by and propagating outwards along the streamer plasma
sheet structure in the wake of the CME-caused streamer
deflections (\textit{e.g.}, Hundhausen, Holzer, and Low, 1987;
Sheeley, Hakala, and Wang, 2000; Tripathi and Raouafi, 2007;
Filippov and Srivastava, 2010). Using the LASCO coronagraph data, Paper
\Rmnum{1} obtains the wave properties like the period, wavelength, and
propagation phase speed for the streamer wave event on 6 July
2004. In the follow-up study by Chen \textit{et al}. (2011, Paper
\Rmnum{2}), a coronal seismological method was developed to
diagnose the values of the Alfv\'en speed and magnetic field
strength in the region surrounding the plasma sheet structure,
with the assumption that the streamer wave is the fast kink
mode propagating along the plasma sheet. Interesting
results concerning the temporal evolution of the physical
conditions of the CME-disturbed corona are found. However, only
one specific streamer wave event which started on 6 July 2004 was
investigated in these studies.

This gives rise to natural questions asking about how many
other similar events exist and, if they do, how the waves get
excited and how their properties compare with the well-studied
2004 event. To address these questions, a complete data survey of
LASCO observations during Solar Cycle 23 was
conducted. It was found that in only about 8 cases there exist
well-defined wavy motions along the streamer stalk among
innumerable CME-streamer interaction events. Therefore, it is
apparent that the generation of streamer waves requires certain
strict excitation conditions to be satisfied. The conditions may
relate to the large-scale coronal magnetic field topology, CME
dynamics and morphological evolution, and physical details of the
CME impact on the streamer structure. One main purpose of this
paper is to provide more clues on the wave excitation conditions
by studying the 8 candidate streamer wave events
observed in Solar Cycle 23 and collecting common observational
features of these events.

The paper is organized as follows. In Section 2, we first present
observations of the 8 events and discuss the physical factors that may
play a role in the wave formation. In Section 3, we conduct further detailed
investigations on a specific event on 5 June 2003 to deduce
the evolution of the wavelength, period, and propagation phase
speed, we also present a coronal seismological study to diagnose
the distribution of the Alfv\'en speed and magnetic field strength
in the plasma sheet region. A summary is provided in the final
section of this paper.

\section{Candidate streamer wave events observed in Solar Cycle 23}
      \label{S-general}
CDAW (Coordinated Data Analysis Workshops) data center provides a
detailed catalog of the CME events observed by LASCO
(Gopalswamy \textit{et al}., 2009), including the associated white light
and running difference images (RDIs) as well as some deduced
physical parameters. Our data survey focusing on the
streamer wave events was conducted taking advantage of the
information gathered by the data center. All CME events in the
CDAW data base during Solar Cycle 23 were viewed to find those in which
the streamer stalk presented snakelike wavy motions
after being hit by a nearby CME. The RDIs were further
examined to confirm the presence of one to a few pairs of
bright-dark and dark-bright (BD-DB) patches along the streamer
stalk, which could be a manifestation of the wavy
motions. Eight candidate streamer wave events, including the
event on 6 July 2004, studied previously, were found. In
the following text, the events will be named after the date on
which the associated CME is first observed. For example, the
just-mentioned event will be called the 20040706 event. Some
relevant physical parameters of the CMEs are presented in the
first to sixth columns of Table 1, including the appearance time
(UT) of the CME ejecta in the C2 field of view (FOV), the CME
type, the central position angle (CPA), the linear speed, the
class of the accompanying flares according to the X-ray flux
recorded by the GOES satellite. The last column of Table 1 gives
the CPA of the associated streamer structure measured at 5 R$_\odot$. In 3
events, it is difficult to rule out the possibility of the wavy
motions being humps pushed by the CME trailing material or magnetic
structures adjacent to the streamer. That is to say in these events it
is possible that both processes, including the action of the magnetic
restoring force as a result of the streamer deflection and the direct
interaction with nearby structures trailing the CME, may play a role in forming the
streamer wavy motions. The 3 events have been indicated by the
symbol ``$\ast$'' after the observation dates given in the first
column.

From the table, we see that most events take place in 2003 and
2004, years of high level of solar activity, and few events occur
in the rising or declining phase of solar activity. This is easy
to understand since statistically CMEs and streamers interact much
more frequently and energetically during times of higher than
lower solar activity. We also find that there are 6 halo CMEs, in
all except one event the apparent angular widths exceed 100$^{\circ}$, and
the average width is about 290$^{\circ}$. The lowest linear speed
is 964 km s$^{-1}$ for the 20030527 event and for all the other
events it is larger than 1000 km s$^{-1}$ with the largest speed
being 2861 km s$^{-1}$ for the 20050115 event, and the average speed
being 1580 km s$^{-1}$. In addition, accompanying flares are
observed for the four front-side events. The flare class varies
from C8.8 to X3.6. The other four events, indicated by the symbol
``$\setminus$'', are mostly back-side events according
to EIT observations (Delaboudini\`{e}re
\textit{et al}., 1995). Considering that all front-side events
have accompanying flares, it is highly possible that this may also
be the case for the back-side CMEs. In the following section, we
will continue to describe some of these observational features and
discuss their relevance to the excitation conditions of streamer
waves.

\subsection{Individual events}
Among the candidate streamer wave events listed in Table
1, events 20040705 and 20040706 have been already reported with
the latter carefully studied in Paper \Rmnum{1} and \Rmnum{2}.
In the following text, we will first present a brief introduction
to these two events, and then describe relevant observations of
the other events in the order of their occurrence.

\subsubsection{Events 20040706 and 20040705 }
According to Table 1, the 20040706 event results from the
interaction between a fast brightness-asymmetric (BA) halo CME
with a linear speed of 1307 km s$^{-1}$ and a streamer structure
with a CPA of 225$^\circ$ as measured at 5 R$_\odot$. The interaction
starts at about 20:30 UT as recorded by C2, causing an obvious deflection of the
streamer structure away from its original position.
The deflection is seen at the bottom of the C2 FOV indicating
that the interaction takes place at an even lower height. At 20:58
UT, the CME ejecta already left the C2 FOV, and the streamer
starts to move backwards. In Paper \Rmnum{1},
it was suggested that the wavy motion is controlled by the magnetic
restoring force given by the deflection of the streamer
structure. The wavy motion is indicated by the BD-DB pairs observed in the
RDIs. The presence of these difference structures facilitates greatly the extraction
of the wave profiles and further quantitative measurements of the
heliocentric distances of selected wave phases, which are given by
subsequent wave crests and troughs observable in the event.

The distances of 5 phases, marked as P1 to P5, are measured
for event 20040706 in Paper \Rmnum{1}. In other words, about two
wavelengths of the streamer wave are observable in the event, with the
distance between P1 to P3 representing the first wavelength and P3 to
P5 the second one. The wavelength is about 2 - 3 R$_\odot$, and the
period is about 1 hour. It is interesting to find that both
parameters increase with the outward propagation of the wave. More
quantitative results and discussions will be presented in Section
3 together with that for event 20030605, to facilitate the
comparison between events. Studies in Paper \Rmnum{1} also
examined the magnetic field topology given by the extrapolation of the
measured photospheric magnetic field with the Potential Field
Source Surface (PFSS) model (Schatten, Wilcox, and Ness, 1969;
Schrijver and De Rosa, 2003). It was concluded that the CME source
lies at the flank of the closed loop systems corresponding to
the streamer. Such geometry makes it possible that the expanding
CME ejecta hits the streamer from the flank, which has been
considered as a physical condition favoring the formation of
streamer wave.

The other event 20040705 results from the interaction between the same
streamer structure and a previous fast halo CME. The CME, taking place,
23 hours before, possibly originates from the same source
region and has a linear speed of 1444 km s$^{-1}$. The RDIs of
this event have been shown in Paper \Rmnum{1} and will not be
presented here. The morphological evolution of the two events
look similar. However, from the RDIs we find that there
exist loop-like trailing eruptive structures moving together with
the streamer kinks. This makes it difficult for us to rule out the
possibility that the wavy motions are streamer humps pushed by the
nearby CME structures. In the following, we shall discuss
relevant observations of the other events listed in Table 1. For
each event, we will pick out several white light images obtained
by C2 to illustrate the CME-streamer interaction process and the
morphology of the streamer wavy motion. For some events, we will
also show several RDIs for better visualizations of the motion.

\subsubsection{Event 20010420}
The CME ejecta appears in the C2 FOV at 10:06 UT with a CPA of
65$^\circ$, an angular width of 127$^{\circ}$, and a linear speed
of 1160 km s$^{-1}$. The CME hits a nearby streamer structure
with a CPA of about 15$^{\circ}$. From the first two white light images
of Figure 1, we observe the deviation of the streamer structure
away from its previous position. This results in the left
(east)-dark and right (west)-bright feature appearing at the
bottom of the first RDI. At 11:30 UT, we see that the CME front
has left the C2 FOV; till 13:31 UT, the streamer stalk presents a
wave-like motion. The wavy motion can be easily related to the
BD-DB pairs in the RDIs of the corresponding intervals. It
should be noted that trailing material continues to flow outwards
even after the CME left, as seen from the corresponding white
light and difference images. The motion of this material
produces the complex difference structures below the CME ejecta.
Some of them are rather close to the streamer structure, and thus
may get mixed up with the bright-dark pairs associated with the
streamer wavy motion, which makes our judgement of the driving
mechanisms of the wavy motions nontrivial. However, with a careful
examination of the white light images, we prefer the explanation
that the motions are driven by the streamer inherent magnetic
restoring force, as an aftermath of the streamer deflection caused
by the CME.

Examining the RDIs, we find that a left-bright and right-dark
feature start to emerge from the bottom of the C2 FOV between 10:54 and
11:06 UT. This color pattern is distinct from
what emerged earlier in association with the CME-streamer
deflection, indicating that the streamer starts to swing backwards
(eastwards), probably under the action of the magnetic restoring force. From
the 11:06 - 11:54 UT RDIs shown in Figures 1k and 1l, we see that the
streamer starts to move westwards again. In the meantime, the
left-bright and right-dark feature observed in Figure 1j already
moves upwards to the location marked by the plus sign, which gives
the first wave crest of the streamer wave. During the interval of
12:06 - 12:30 UT, we observe again the eastward motion of the streamer
structure. Therefore, the streamer structure completes one period
of the wavy motion from 11:06 UT to 12:30 UT. The period is thus
estimated to be about 1 - 1.5 hours. In the last three RDIs, the
locations of the following wave trough and crest are marked by the
stars and squares. It is roughly estimated that the wavelength
of the streamer wave is about 2 - 3 R$_\odot$ and the phase speed
is about 300 - 400 km s$^{-1}$.

\subsubsection{Event 20030527}
At 23:50 UT on May 27, LASCO recorded a halo CME with a linear
speed of 964 km s$^{-1}$ accompanied by an X1.5 solar flare. At
00:06 UT on the following day, another BA halo CME with a linear
speed of 1366 km s$^{-1}$ accompanied by an X3.6 flare is
observed. According to the corresponding EIT data, we find that
both CMEs probably originate from the same active region.
During the eruptions, both CMEs trigger wavy motions of a nearby
streamer with a CPA being about 185$^{\circ}$, as shown in Figure 2. The
most interesting observational feature seen from this figure is
that the streamer exhibits a waving and tangling morphology from
00:50 to 01:50 UT. Possible reasons accounting for the tangling
morphology can be deduced by examining the Wilcox Solar Observatory Source Surface
Synoptic Charts (http://wso.stanford.edu/synsourcel.html) for
CR2003. It is found that the current sheet structure relevant to
the streamer in question lies in between 50$^{\circ}$ -
60$^{\circ}$ southern latitude and extends over a rather wide
longitudinal range from 200$^{\circ}$ - 300$^{\circ}$.
Thus, it is suggested that distinctive wavy motions can be supported by
different longitudinal parts of the plasma sheet, whose
projections onto the sky plane form the observed tangling
morphology. During the event, the CME structures associated with
both CMEs flow outwards persistently, and we are unable to
tell the exact mechanism driving the wavy motions.

\subsubsection{Event 20030605}
From Table 1, we see that the CME appears in the C2 FOV at 20:06
UT with a CPA of 230$^{\circ}$, an angular width of 310$^{\circ}$,
and a linear speed of 1458 km s$^{-1}$. At 20:30 UT, the CME hits a
streamer structure with a CPA being about 316$^{\circ}$ from the bottom
of the C2 FOV, as seen from Figure 3a. This results in an obvious leftward
deflection, and a subsequent rightward motion of the streamer structure.
It can be seen that the CME front leaves the C2 FOV very rapidly within
about 1.5 hours after its first appearance. The majority of the
CME ejecta is present on the right side of the streamer. On the other side,
there exist structures with relatively weak brightness moving outwards along
with the CME front. The material trailing these weak frontal structures
are looking faint from the white light observations on the left side of the
streamer. We therefore take this event as a streamer wave event suggesting
that the wavy motion is supported by the inherent magnetic restoring force
of the deflected structure. The morphology of the wavy motion can be observed from
the last two images taken at 21:30 UT and 21:54 UT, when most part of
the CME already leaves the C2 FOV. It is found that the event is
simple and clear with more than three wave phases
observable. This allows us to conduct a further investigation
on the wave properties, which will be done as we proceed.

\subsubsection{Event 20031118}
The CME driving this event appears in the C2 FOV at 8:50 UT with
an accompanying M3.9 flare and a linear speed of 1660 km s$^{-1}$,
whose image observed at 9:06 UT is given in Figure 4a. At 10:26 UT,
another CME with a CPA of 95$^{\circ}$ and a linear speed of 1824
km s$^{-1}$ is observed by C2 and shown in Figure 4b. The ejecta
of the first CME interacts with a streamer structure with a CPA of
about 319$^{\circ}$. From Figure 4a, we see that the interaction
starts from the bottom of the C2 FOV or probably at an even lower height, which
results in a leftward deflection of the streamer structure. The
RDIs in the streamer region rotated by 41$^{\circ}$
counterclockwise are shown in Figures 4c - 4f. From Figure
4c, it is easy to see that the CME-caused deflection,
corresponding to the left-bright and right-dark feature in the
upper part of the figure, is followed by a rightward motion as
indicated by the right-bright and left-dark feature in the lower
part, while no obvious trailing material of the CME
is present on the left side of the
streamer. Therefore, it is suggested that the above rightward
motion is supported by the inherent magnetic restoring force of
the streamer, and thus gives another example of streamer wave
event.

According to the timing and the enhanced deflection of the
streamer structure as seen from the second right-bright and left-dark
feature at the bottom of Figures 4e and 4f, we suggest that this
feature represents the streamer deflection caused by the second
CME. It is found that the associated ejecta has no direct contact
with the streamer structure. So it is possible that
the impact of the CME on the streamer is achieved by expelling the surrounding coronal
magnetic field and/or propagating disturbances. Examining the data
from Wind/WAVES (Bougeret, \textit{et al}., 1995), we find that
there exists a type \Rmnum{2} radio burst at 10:10 UT, corresponding
temporally to the presence of the lower right-bright and left-dark
feature of Figure 4e. Therefore, it is possible that
the feature is a result of the streamer deflection caused by the associated
CME shock (see, \textit{e.g.}, Sheeley, Hakala, and Wang, 2000). The subsequent observation
indicates that this deflection does not grow into a streamer wave,
yet terminates the development of the wave event caused by the
previous eruption.

\subsubsection{Event 20050115}
At 23:06 UT, a halo CME appears in the C2 FOV with a linear speed
as large as 2860 km s$^{-1}$ and an accompanying X2.6 solar flare.
According to Figures 5a and 5b, we see that the eruption results
in a global perturbation with all the streamers observed in the plane
 of the sky deflected. Among the streamers, the one with a
CPA of 240$^{\circ}$ at 5 R$_\odot$ presents the largest amplitude of deflection,
along which the concerned wavy motion develops. In
Figures 5c - 5f, we show the corresponding RDIs in the streamer
region within the range of 4 R$_\odot$ to 12 R$_\odot$ taken from
the C3 observations. The right-bright and left-dark feature in the
upper part, \textit{i.e.}, father away from the Sun, of Figure 5c is associated with the rightward
deflection of the streamer structure in response to the CME
impact. The feature with different brightness distribution
in the lower part of this figure is believed to be caused by the
inherent magnetic restoring force of the streamer since no obvious
trailing structures are observed on the right side of the streamer. In
the last three panels of RDIs, we observe the outward propagation
of the bouncing motion. At the lower part of Figures 5e and 5f we can
discern the presence of another rightward wavy motion of the
streamer structure. Thus, only one pair of the BD-DB feature is
observable in this event.

\subsubsection{Event 20061106}
The CME driving this wave event is present in the C2 FOV at 17:54 UT
with a CPA of about 80$^{\circ}$, an angular width of about
80$^{\circ}$, and a linear speed of 1994 km s$^{-1}$. The streamer
with a CPA of about 128$^{\circ}$ is deflected by the ejecta, as seen
from Figures 6a and 6b. From Figures 6b and 6d, we see that the
CME front already leaves the C2 FOV at 18:54 UT. The streamer starts
to wave rightwards from the CME deflection producing the left-dark
and right-bright feature in the streamer region at the bottom of
Figure 6d. From Figures 6e and 6f, we see that the feature
continues to propagate outwards along the streamer stalk to the
outer edge of the FOV. There are no obvious trailing structures
observed on the left side of the streamer. So this event is
 suggested to be a streamer wave event. It is also found
that there is no direct contact of the CME ejecta with the
streamer structure indicating that the interaction is possibly achieved by
the CME expelling the surrounding coronal magnetic field and/or
relevant disturbances. The Wind/WAVES data show that there exists
a type \Rmnum{2} decametric radio burst at about 18:00 UT. Therefore,
it is also possible that the streamer wave is related to the shock
disturbance driven by the eruption.

\subsection{Discussion on the excitation conditions of streamer waves}
  \label{S-labels}

As mentioned previously, the excitation or formation of streamer
wave requires certain strict physical conditions to be satisfied.
The conditions may be related to the large-scale magnetic geometry
of the corona, the CME dynamics and morphology, as well as the
details of the CME-streamer interaction. To provide clues on these
conditions, we first summarize the common observational features
among the candidate events of streamer waves discussed in
the previous subsection. First, all the driving CMEs are fast and
wide with an average linear speed of 1580 km s$^{-1}$ and an
average apparent angular width of 290$^{\circ}$. Second, all the front-side
CMEs have accompanying flares indicating the occurrence of
magnetic reconnections during the process. Third, in most events
CMEs hit the streamer from the flank, and the sites where the
impact takes place are no higher than the bottom of the C2 FOV,
\textit{i.e.}, lower than about 2 R$_{\odot}$ in heliocentric distance.
Keeping these observational common characteristics in mind, in the
following we explain why the above conditions can be taken as
necessary conditions for the excitation of streamer waves. We
start by discussing the physical origin of the associated
restoring force.

It is generally believed that the dynamical equilibrium between
the expansion of hot coronal plasmas and the confinement of closed
magnetic arcades gives rise to coronal streamers consisting of
closed field arcades rooted on the photosphere and a cusp atop of
them. Upon the impact of the CME ejecta, the streamer structure
below the cusp gets deflected away from its equilibrium
position. Due to the photospheric line-tying effect of the
deflected field lines, the streamer structure responds to the CME
impact with a magnetic restoring force, which may be given by both
the magnetic tension and pressure. Thus, to produce such a
restoring force, the closed field arcades in the streamer should
be deflected. This requires that the site where the initial
interaction between the ejecta and the streamer structure takes
place be low enough, in agreement with the
observations. Furthermore, to have a restoring force strong enough
to produce observable effects on the streamer structure, the
following conditions should also be satisfied.

First, the deflection amplitude should be large enough.
This requires that the CME ejecta be close enough to the
streamer, and the CME expands fast enough, also consistent
with the observations of most of our events. Second, the
duration of the CME interacting with the streamer structure below
the cusp, say from 1.5 $-$ 2.5 R$_\odot$, should be short in
comparison with the acting time scale of the inherent restoring
force. For a fast-moving CME with a speed no less than 1000 km
s$^{-1}$, the ejecta can travel more than 5 R$_\odot$ in 1 hour.
In this case, the duration of the CME-streamer interaction in the
region between about 1.5 $-$ 2.5 R$_\odot$ is appreciably short
compared to the time scale of the restoring force which is
estimated to be about 1 hour according to our observations.
Therefore, a fast CME is also believed to be one necessary
excitation condition for the growth of streamer waves. In
addition, to make space for the bouncing motion of the streamer
structure following the deflection, some of the magnetic field
lines stretched outward by the ejecta should close back. This
requires the occurrence of magnetic reconnections along with the
eruptions, consistent with the presence of solar flares with all
front-side CMEs.

We note that the above conditions are only necessary ones,
other conditions including the details of the
CME-streamer interacting process may be also required. On the
other hand, it is true that whether the streamer wave, if already
formed, is observable with the coronagraph still depends on the
observational view angle, presence of interfering bright
structures in the fore- and background corona, and existence of
disturbing features of the present eruption and/or other nearby
CMEs. We note that the work on excitation conditions of
streamer waves is expected to improve by the analysis of more
events in the future, as well as magnetohydrodynamic
(MHD) modeling efforts designed to study the physics of
CME-streamer interaction.

\section{Wave properties and relevant seismological study on the 20030605 event}
      \label{S-features}
This is suggested to be a simple and clear streamer wave event,
which allows us to conduct a further analysis of the wave
properties. In the following subsection, we will first determine
the heliocentric distances of four wave crests/troughs based
on the LASCO observations; then, we deduce the radial profiles of
the wave period and wavelength, as well as the
propagating phase speeds. In the second subsection, following the
seismological approach developed by Paper \Rmnum{2} we estimate
the radial profiles of the Alfv\'en speed and magnetic field
strength in the region surrounding the plasma sheet. The results
will be compared with the 20040706 event studied previously.

\subsection{Wave properties}
  \label{S-equations}
Four white light images of this event have been presented in
Figure 3. In Figure 7, we show all the relevant RDIs given by
LASCO C2 (lower panels) and C3 (upper) observations. From the
RDIs, it is easy to recognize the dark-bright pairs corresponding
to the streamer deflection and the subsequent wavy motion.
Using these difference images, we delineate
the profiles of the streamer wave, and determine four specific
wave phases at which the wavy motion reaches the maximum amplitude.
The phases are called as P1, P2,
P3, and P4, and marked with plus signs, asterisks, squares, and
triangles. In comparison with the first three phases, P4 appears
later associated with a smaller wave amplitude and a lower number
of observations and less accuracy of distance and speed
measurements as a result. Therefore, in the following text we will
focus on the behavior of the first three phases.

The variations of the heliocentric distances of the four phases are
shown in Figure 8a with corresponding symbols. The solid lines are
given by the second order polynomial fittings to the measurements,
which are then used to deduce the radial and temporal profiles of
the propagation speed of a certain phase. Before presenting the
speed profiles, we first discuss the variation of the wave period
and wavelength deduced from Figure 8a.

For streamer waves, the wavelength is defined as the distance
between two adjacent wave crests or wave troughs, \textit{e.g.}, the
distance between P1 and P3 at a fixed time; the
wave period is defined as the time difference of P1 and P3 passing
the same altitude. With these definitions, radial evolutions of the two parameters can be
deduced from Figure 8a and are shown in Figures 8c and 8d, respectively.
Results for the 20040706 event are also plotted with dashed lines.
We can see that, for the 20030605 (20040706)
event, the period and wavelength increase from 2.4 R$_\odot$ and
70 minutes (2.0 R$_\odot$ and 50 minutes) at 4 R$_\odot$ to 2.8
R$_\odot$ and 85 minutes (2.6 R$_\odot$ and 85 minutes) at 8
R$_\odot$. Both parameters exhibit similar increasing trend
with distances. It should be pointed out
that, the statement regarding the wavelength increasing with
distance has been explained with the positive
difference of speeds of preceding and trailing wave phases in Paper \Rmnum{1}.
For example, P1 moves faster than P2 and P3, as observed, this
makes the distance between P1 and P3, \textit{i.e.}, the first wavelength,
increases with the wave propagation. Nevertheless, the increase of
period with the wave propagation is not mentioned in previous
studies. Here, we provide an explanation of this observation still
making use of the difference of phase speeds. From the definition of
wave period given above, the time required by P1 to pass through a certain
distance range, say from 4 to 6 R$_\odot$, is shorter than the
time used by P3 since at a fixed distance P1 moves faster than P3  in general.
It can be found that in this case the wave period increases with distance.
The phase speed variations involved here have
both the spatial and temporal contributions, which are likely associated
with the recovering process of the CME-disturbed corona. We will
further discuss this process in the following subsection.

The phase speeds deduced from the distance-time fittings are
plotted as solid lines in Figure 8b. From this figure, we see that
the speeds of both P1 and P2 decrease from 435 and 400 km s$^{-1}$
at 3 R$_\odot$ to 425 and 390 (415 and 382) km s$^{-1}$ at 5 (7)
R$_\odot$, while the speed of P3 keeps almost constant at about
371 km s$^{-1}$. In addition, the speed averages of the
four phases are 402, 376, 371, and 350 km s$^{-1}$, which
basically decrease with the order of their presence.
Comparing the results of events 20030605 and 20040706, we
find that the values and variation trends of the speeds
associated with individual phases are essentially similar.

\subsection{Relevant seismological study}\label{S-simple-equations}
Using measurements of the 20040706 event, Paper \Rmnum{2}
conducted a seismological study to diagnose the distribution of
the Alfv\'en speed and magnetic field strength in the region
surrounding the streamer plasma sheet, interpreting the streamer
wave as the fast kink body mode propagating along the sheet
structure. The measured phase speed $v_p$ has two contributions,
one is the speed of the background solar wind $v_{sw}$, the other
is the phase speed of the mode at the plasma rest frame $v_k$.
The possible variation range of the former parameter is constrained
by the statistical results of the speeds of blobs flowing
together with the wind along the plasma sheet (Wang \textit{et al}., 2000).
The latter is connected to the Alfv\'en speed in the region
outside of the plasma sheet ($v_{Ae}$) with an approximation
deduced from a parameter study of the corresponding dispersion
relation given by Edwin and Roberts (1982). We note that
the situation considered by them, where a slab
is embedded in an otherwise uniform environment, is different from
the present case where an electric current sheet exists inside the
high density slab, as suggested by the PFSS extrapolation results
for all streamers studied here. The magnetic field reverses its
direction across this sheet, thereby presenting a further
transverse structuring. However, it can be shown that this current
sheet is transparent to the two-dimensional perturbations we
consider, as long as it is infinitely thin. Details of our
deductions are given in the Appendix.

Thus, the radial profiles of $v_{Ae}$ can be deduced for
individual phases. With the electron density distribution limited
by the inversion of the pB (polarized brightness) data (van de Hulst,
1950; Hayes, Vourlidas, and Howard, 2001) recorded by LASCO, the magnetic field
strengths associated with individual phases are then evaluated.
It should be pointed out that these diagnoses have large
errors and uncertainties. Measurement errors of the phase speeds
are estimated to be about $\pm$10$\%$ (or a total of 20$\%$) in
Paper \Rmnum{1}, and factors contributing to the uncertainties of
our seismological studies are discussed in Paper \Rmnum{2} and
will not be repeated here.

Following the approach developed in Paper \Rmnum{2}, we conduct a
seismological study using the streamer wave observations in the
20030605 event. For simplicity, only the average speed of the blob
measurements is used in the study, which is plotted as the
dot-dashed line in Figure 9a. It can be seen that the solar wind
speed increases gradually from 50 km s$^{-1}$ at 3 R$_\odot$ to
110 (150) km s$^{-1}$ at 5 (7) R$_\odot$. Subtracting the above
values of $v_{sw}$, we get the phase speed of the kink mode at the
plasma rest frame $v_k$. According to the parameter study on the
dispersion relation carried out in Paper \Rmnum{2}, $v_k$ is related
to $v_{Ae}$ with $v_k \approx \alpha v_{Ae}$, where $\alpha
\approx 0.92$. Note that this value of $\alpha$ is determined with
the electron density distribution for the 20040706 event, which is
basically similar to that for the present event as seen from the
results of the pB inversion plotted in the following figure.
Therefore, we make use of the same value of $\alpha$ in this
study.

The obtained radial profiles of $v_{Ae}$ associated with phases
P1, P2, and P3 are shown in Figure 9a with solid, dotted, and
dashed lines, respectively.  From this figure, we see that values
of $v_{Ae}$ are 408 (P1), 373 (P2), and 343 (P3) km s$^{-1}$ at 3
R$_\odot$, and decrease to 341 (P1), 304 (P2), and 283 (P3) km
s$^{-1}$ at 5 R$_\odot$ and 287 (P1), 249 (P2), and 237 (P3) km
s$^{-1}$ at 7 R$_\odot$. It can also be seen that in the region
surrounding the plasma sheet the Alfv\'enic critical point,
where $v_{Ae}=v_{sw}$, lies at $\sim$ 10 R$_\odot$ and the
Alfv\'en speed is $\sim$ 200 km s$^{-1}$ there. This result is in
agreement with the previous deduction using data from the Helios
spacecraft (Pizzo \textit{et al}., 1983).

The radial profiles of the electron density below 5 R$_\odot$ can
be obtained with the pB inversion method given in the
SolarSoft package, and that beyond 5 R$_\odot$ are given by
assuming an $r^{-2}$ dependence. The pB data on 5 June 2003 were
recorded at 21:05 UT when the streamer exhibits significant
snakelike motion and, therefore, are not
appropriate for the density inversion along a specific radial
direction. There are no data obtained on the previous day, so
we make use of the pB data recorded at 21:00 on 6 June for the
required density deduction. This is equivalent to assume that
the density along the streamer structure does not change
appreciably during the day after the CME eruption. The densities
along position angles of 315$^\circ$ and 300$^\circ$, representing
different directions along and away from the streamer, are
plotted in Figure 10b with dotted and dashed lines, respectively.
The average of the two sets of densities is delineated with the
solid line with values of $6.5 \times 10^{5}$ $(8.6 \times
10^{4}$, $4.4 \times 10^{4})$ cm$^{-3}$ at 3 (5, 7) R$_\odot$,
this is substituted in the deduced profiles of $v_{Ae}$ to
diagnose the magnetic field strengths $B_e$ in the corresponding
region. The obtained field strengths are shown in Figure 9b for
phases P1 (red-solid), P2 (blue-dotted), and P3 (yellow-dashed).
From the figure, we see that $B_e$ is 0.145 (0.045, 0.028) G for
P1, 0.139 (0.041, 0.024) G for P2, and 0.129 (0.038, 0.023) G for
P3 at 3 (5, 7) R$_\odot$, respectively. Basically $B_e$ decreases
according to the $r^{-2}$ dependence.

In Figure 9b, we also show other estimates on the magnetic field
strength in the corona with various symbols, including results
from our previous study employing the streamer wave method to the
20040706 event presented as shadow areas, and
several other diagnoses obtained employing various radio methods.
To be specific, the strength-distance relationship in the
heliocentric range of 1.02 - 10 $R_\odot$ above active regions
(dot-dashed line) given by Dulk and McLean (1978) is mainly
based on radio burst observations, the results of Vr\v{s}nak \textit{et al}.
(2002) and Cho \textit{et al}. (2007) presented as crosses and diamonds are
deduced using the band-splitting phenomenon of type \Rmnum{2} radio
bursts, the results from the Faraday-rotation measurement of radio
signals emitted from the Helios spacecraft and extragalactic radio
sources are included as open circles with error bars (P\"{a}tzold
\textit{et al}., 1987), triangles (Spangler, 2005), and squares (Ingleby \textit{et
al}., 2007). The latest results obtained by Ramesh \textit{et al}. (2010)
employing the low-frequency circularly polarized radio emission
inside a streamer structure are given as solid inverse triangles.
We note that the above list of diagnoses of the coronal field
strength is incomplete, and there exist many other estimates (see
\textit{e.g.}, references in Vr\v{s}nak \textit{et al}., 2002). From this figure, it can
be seen that the magnitude and variation trend of the magnetic
field strength in the region surrounding the plasma sheet are
basically consistent with the results for event 20040706 and
other estimates.

In addition, we also find that both the Alfv\'en speed and
magnetic field strength have a general decline trend with
time at a fixed distance. According to the explanation provided
in Paper \Rmnum{2}, this trend of variation of the two quantities
is a result of the recovering process of the CME-disturbed corona.
From Figure 9, it can be seen that both the magnetic field
strength and Alfv\'en speed decline by about 15\% during about
60-90 minutes, \textit{i.e.}, in about one wave period. Of course, the
coronal field can not decrease unlimitedly. However, at this time
it is not possible to assess how the magnetic field strength may
evolve in such a short interval in the absence of a well-observed
streamer wave phenomenon. Finally, we note that the
differences between the diagnostic results for the three wave
phases are possibly not significant considering the errors and
uncertainties associated with our diagnostics.

\section{Conclusions}

In this paper, we present 8 candidate streamer wave
events found via a data survey through LASCO observations in
Solar Cycle 23. We find the following three common
observational features in these events: (1) the driving CMEs
are wide and fast with a linear speed no less than $\sim$ 1000 km
s$^{-1}$; (2) all front-side CMEs have accompanying flares; (3) in most events the
bright CME and streamer structures have direct contact with each
other, and the interaction starts at a height no higher than the
bottom of the C2 FOV, \textit{i.e.}, lower than $\sim$ 2 R$_\odot$. These
common features shed light on the excitation conditions of
streamer waves. Nevertheless, more similar events, when available
in the future, should be analyzed and MHD models should be developed
for a better understanding on the
physics of CME-streamer interaction.

A further study on the event dated on 5 June 2003 gives radial and
temporal evolution of wave properties including the period,
wavelength, and propagation speeds of four observable wave
crests/troughs. It is found that both the period and wavelength
increase gradually with the wave propagation along the streamer
plasma sheet, and the phase speed of the preceding wave phase is
generally faster than that of the trailing ones. The associated
coronal seismological study yields the radial profiles of the
Alfv\'en speed and magnetic field strength in the region
surrounding the streamer plasma sheet. It is found that the
Alfv\'enic critical point is at about 10 R$_\odot$ where the flow
speed, which equals the Alfv\'en speed, is $\sim$ 200 km s$^{-1}$.
The magnetic field strengths corresponding to the first three wave
phases are 0.145, 0.139, and 0.129 G at 3 R$_\odot$, and decrease
generally according to the $r^{-2}$ dependence to 0.045, 0.041,
and 0.038 G at 5 R$_\odot$, and to 0.028, 0.024, and 0.023 G at 7
R$_\odot$, respectively. The obtained results are generally
consistent with that of another well-studied event on 6 July 2004.

\begin{acks}
The SOHO/LASCO data used here are produced by a consortium of the
Naval Research Laboratory (USA), Max-Planck-Institut f\"{u}r
Aeronomie (Germany), Laboratoire d'Astronomie Spatiale (France),
and the University of Birmingham (UK). SOHO is a project of
international cooperation between ESA and NASA. We thank Dr. A.
Vourlidas for helping us analyze the LASCO pB data. This work was
supported by grants NNSFC 40825014, 40890162, A Foundation for the
Author of National Excellent Doctoral Dissertation of PR China
(2007B24), and the Specialized Research Fund for State Key
Laboratory of Space Weather in China. B Li is supported by the
grant NNSFC 40904047, LD Xia by 40974105, and HQ Song by
Natural Science Foundation of Shandong Province ZR2010DQ016
and Independent Innovation Foundation of Shandong University
2010ZRYB001.
\end{acks}

\appendix
We note that the situation considered by Edwin and Roberts (1982, ER82 from now on),
where a slab is embedded in an otherwise uniform environment, is
different from the present case where an electric current
sheet (CS) exists inside the high density slab. The magnetic field
reverses its direction across this sheet, thereby presenting a
further transverse structuring. However, it can be shown that this
CS is transparent to the two-dimensional perturbations we
consider, as long as it is infinitely thin. To illustrate this,
let us start with Figure 1 in ER82 where the
vertical lines $x=\pm x_0$ separate the slab from its environment.
The dispersion relation (Equation (11) in ER82) then
follows from the linearized ideal MHD equations, the ansatz that
any small-amplitude time-dependent perturbation $f(x,z;t)$ is in
the form $\hat{f}(x)\exp[i(\omega t + kz)]$, the condition that
the perturbations are not identically zero, as well as the
requirements that the perturbed total pressure (gas plus magnetic)
and transverse velocity $v_x$ be continuous across the interfaces
$x=\pm x_0$. With the above ansatz in mind, a pair of $(\omega,
k)$ that satisfies the dispersion relation will give a $\hat{v}_x$
continuous throughout the entire $x-z$ plane. To be specific,
Equation (10) in ER82) expresses both kink and
sausage modes, with the former satisfying
$\beta_e=-\alpha_e=\beta_0\sinh(m_0 x_0)$, and the latter obeying
$\beta_e=\alpha_e=\alpha_0\cosh(m_0 x_0)$. Here $\alpha_0$ or
$\beta_0$ is arbitrary. Among the rest of perturbations pertaining
to these modes, $\hat{v}_z$, $\hat{p}_T$ are given by Equations (17) and
(18) in Roberts (1981), while $\hat{b}_x$ and $\hat{b}_z$ can be
found via Equation(10) in Roberts (1981), $\hat{p}$ via Equation (9) and
$\hat{\rho}$ via Equation (6a). Here $\vec{v}=(v_x, 0, v_z)$ and
$\vec{b}=(b_x, 0, b_z)$ are the velocity and magnetic field
perturbations, respectively. Moreover, $p, \rho$ and $p_T$
represent the perturbed pressure, density and total pressure. The
hat $\hat{}$ refers to the Fourier amplitudes. Let the solution
set $(\hat{\vec{v}}, \hat{\rho}, \hat{p}, \hat{\vec{b}})$ be
denoted by $S^{\mathrm{ER82}}$. Now suppose the sign of the
magnetic field in equilibrium in the right half of the plane
($x>0$) turns negative. It turns out that $S^{\mathrm{ER82}}$ is
also a solution of the linearized MHD equations in this case, the
only modification being that $\hat{\vec{b}}$ should be in the
opposite direction to $\hat{\vec{b}}$ found in ER82 for $x>0$.
Across the (perturbed) interface initially
located at $x=0$ in equilibrium, the normal magnetic field,
transverse displacement, and the perturbed total pressure $p_T$
are all continuous. (That $p_T$ is continuous follows from the
fact that $p$ and $B_0 b_z$ are both continuous, and one can
easily see that the latter does not change if $B_0$ and $b_z$
changes their signs simultaneously.) This allows us to see the CS
as transparent within the framework of linear, ideal MHD. In
particular, the dispersion relation derived in ER82 also applies.

\begin{table}[!htbp]
\caption {Some relevant physical parameters of the CMEs and streamers
of the 8 candidate streamer wave events. The first to
sixth columns present the appearance date and time (UT) of the CME
ejecta in the C2 field of view (FOV), the CME type, the central
position angle (CPA), the linear speed, and the importance of the
accompanying flares according to the X-ray flux recorded by the
GOES satellite. The last column gives the CPA, measured at 5 R$_\odot$,
of the streamer structure with wavy motion.}

\begin{tabular}{lcccccc} 
  \toprule
  \multicolumn{6}{c}{CME}    & Streamer                                          \\
   \cline{1-6}
  Date      &Time       &CPA        &Width   &Speed       &    Flares & CPA         \\

 (yyyy/mm/dd)  &(UT)         & (deg)     & (deg)  &(km s$^{-1}$)&     &(deg) \\
 \midrule
 20010420   &10:06:05   &65         &127     &1160        & $\setminus$        &   15      \\
 20030527*   &23:50:05   &Halo/S     &360     &964        &X1.3       &  185     \\
(20030528)  &(00:50:05) &(Halo/BA)  &(360)   &(1366)      &(X3.6)     & (185)    \\
20030605    &20:06:05   &230        &239     &1458        & $\setminus$        &316       \\
20031118*    &08:50:05   &Halo/BA)   &360    &1660       & M3.9      & 319      \\
20040705*    &23:06:05   & Halo     &360     &1444        &  $\setminus$        & 225      \\
20040706    &20:06:06   &Halo(BA)       &360     &1307     &$\setminus$       &225       \\
20050115    &23:06:50   &Halo(BA)       &360     &2861        & X2.6      & 240      \\
20061106    &17:54:04   &80         &80      &1994        &C8.8       &128    \\

  \bottomrule
\end{tabular}
\end{table}
 \begin{figure}    
   \centerline{\includegraphics[width=1\textwidth]{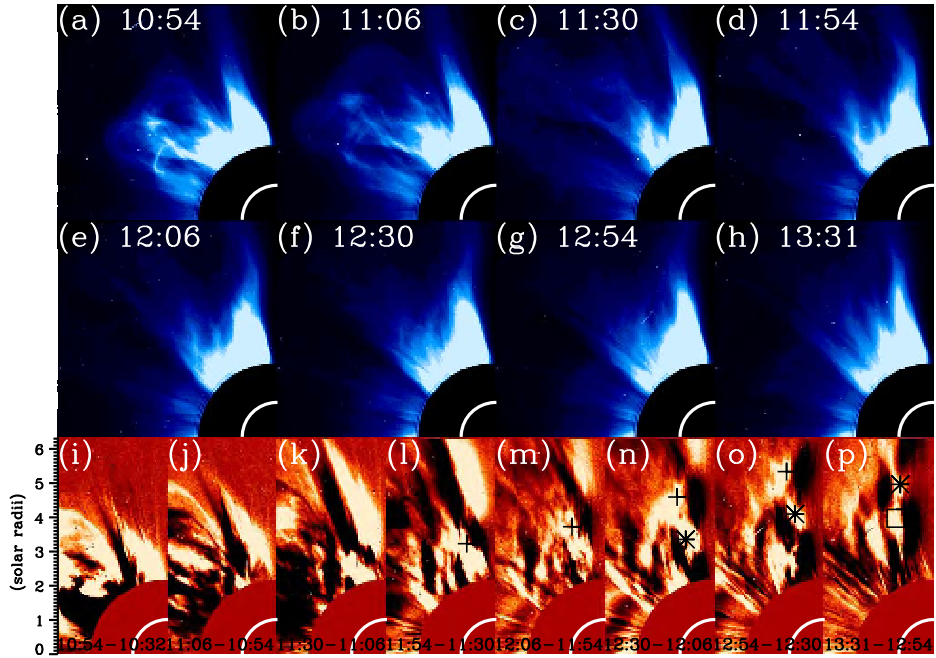}
              }
              \caption{White light images and RDIs in the streamer
              region observed by LASCO C2 for the 20010420 event.
              The plus signs, asterisks, and squares
              mark the location of phases P1, P2, and P3. See text
              for details.}
   \label{f1}
   \end{figure}
\begin{figure}    
   \centerline{\includegraphics[width=1\textwidth]{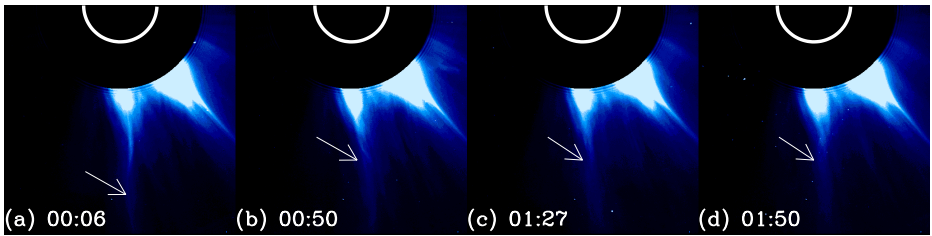}
              }
              \caption{White light images of the disturbed streamer
              structure observed by C2 on 28 May 2003. Arrows
              indicate the waving and tangling features of the
              streamer stalk.
                      }
   \label{f2}
   \end{figure}
\begin{figure}    
   \centerline{\includegraphics[width=1\textwidth]{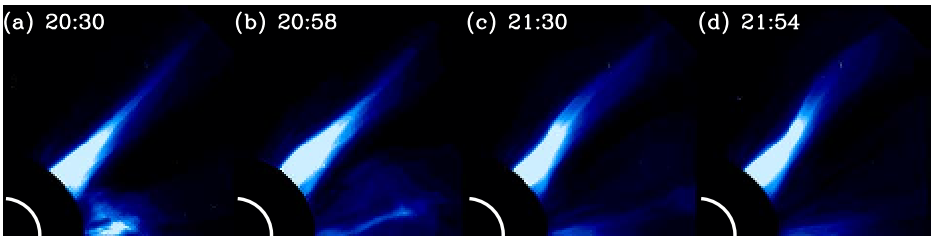}
              }
              \caption{White light images of the CME-streamer interaction event observed by LASCO C2 on 5 June 2003.}

   \label{f3}
   \end{figure}
\begin{figure}    
   \centerline{\includegraphics[width=1\textwidth]{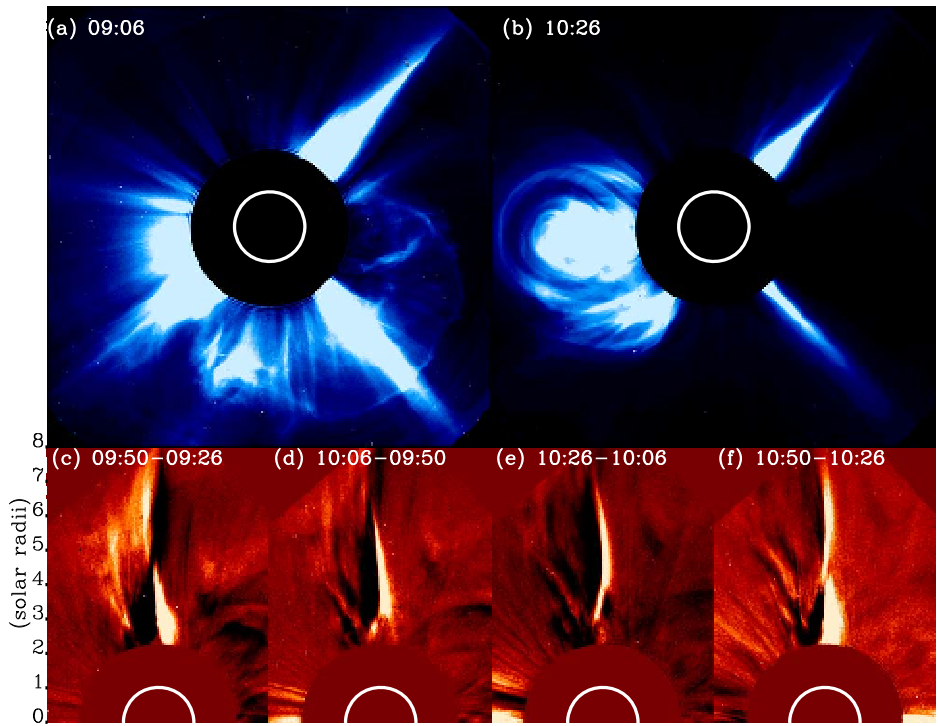}
              }
              \caption{White light images (a, b) of the two CME eruptions,
              and RDIs (c-f) of the streamer region for the 20031118 event.
              The RDIs have been rotated counterclockwise by 41$^{\circ}$.}
   \label{f4}
   \end{figure}
\begin{figure}    
   \centerline{\includegraphics[width=1\textwidth]{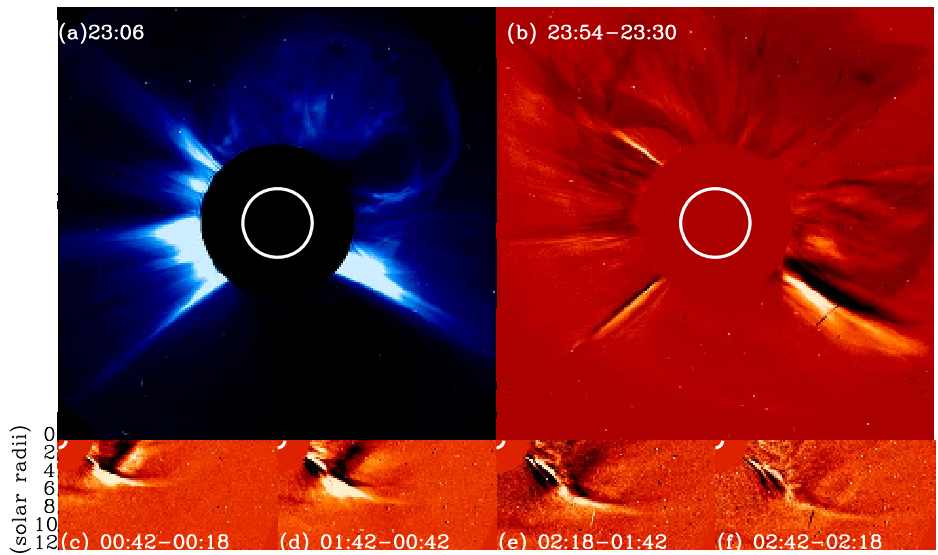}
              }
              \caption{A white light image (a) and RDIs (b-f) for
              the CME-streamer wavy event observed by LASCO C2
              on 15 January 2005. The CPA of the wavy streamer is
              240$^{\circ}$. The RDIs in the lower panels are in the
              range of heliocentric distances from 4 R$_\odot$ to 12 R$_\odot$
              taken from the C3 observations.
                      }
   \label{f5}
   \end{figure}
\begin{figure}    
   \centerline{\includegraphics[width=1\textwidth]{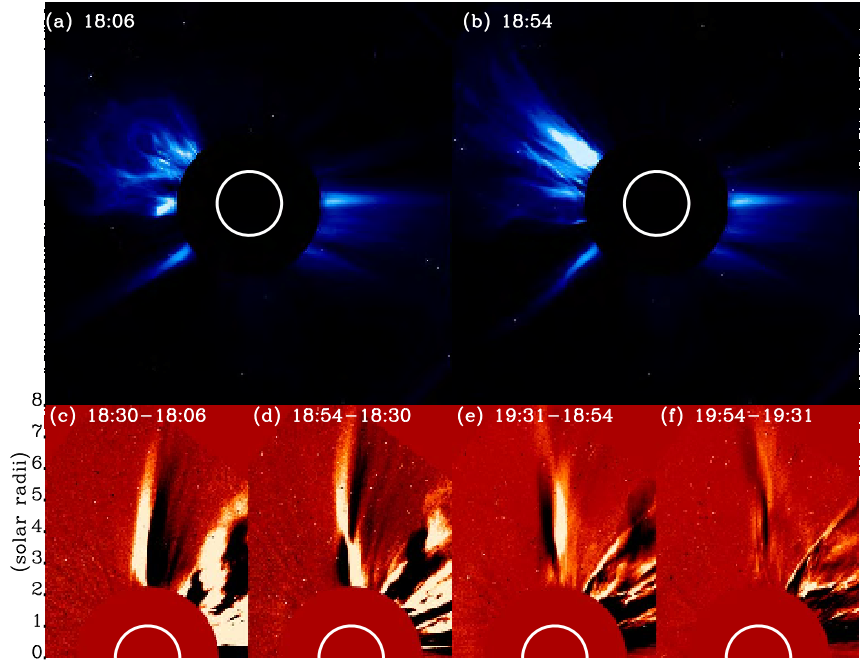}
              }
              \caption{White light images (a, b) and RDIs (e-f) of the
              CME-streamer interaction event observed by LASCO C2 on 6 November 2006.
              The CPA of the wavy streamer is 128$^{\circ}$. The RDIs in the
              lower panels have been rotated clockwise by 128$^{\circ}$.
                      }
   \label{f6}
   \end{figure}
\begin{figure}    
   \centerline{\includegraphics[width=1\textwidth]{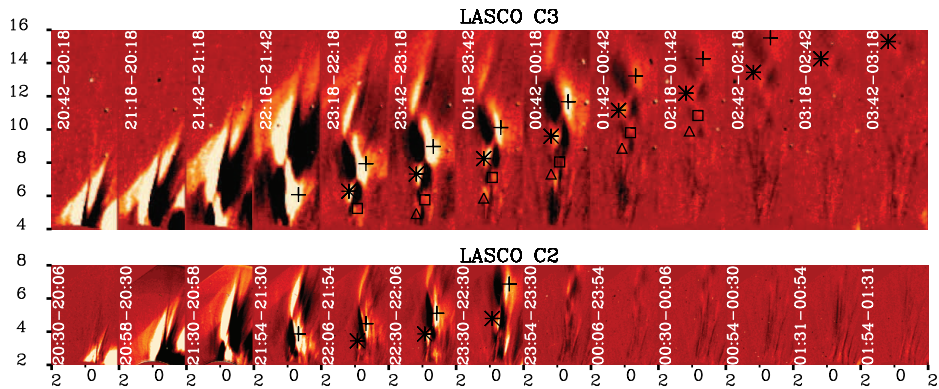}
              }
              \caption{RDIs in the streamer region for the 20030605 event
              observed by LASCO C2 (lower panels) and C3 (upper).
              All images have been rotated counterclockwise by 35$^{\circ}$.
              The four selected wave phases (P1 - P4) are marked by plus signs, asterisks,
              squares, and triangles, respectively.}

   \label{f7}
   \end{figure}
\begin{figure}    
   \centerline{\includegraphics[width=1\textwidth]{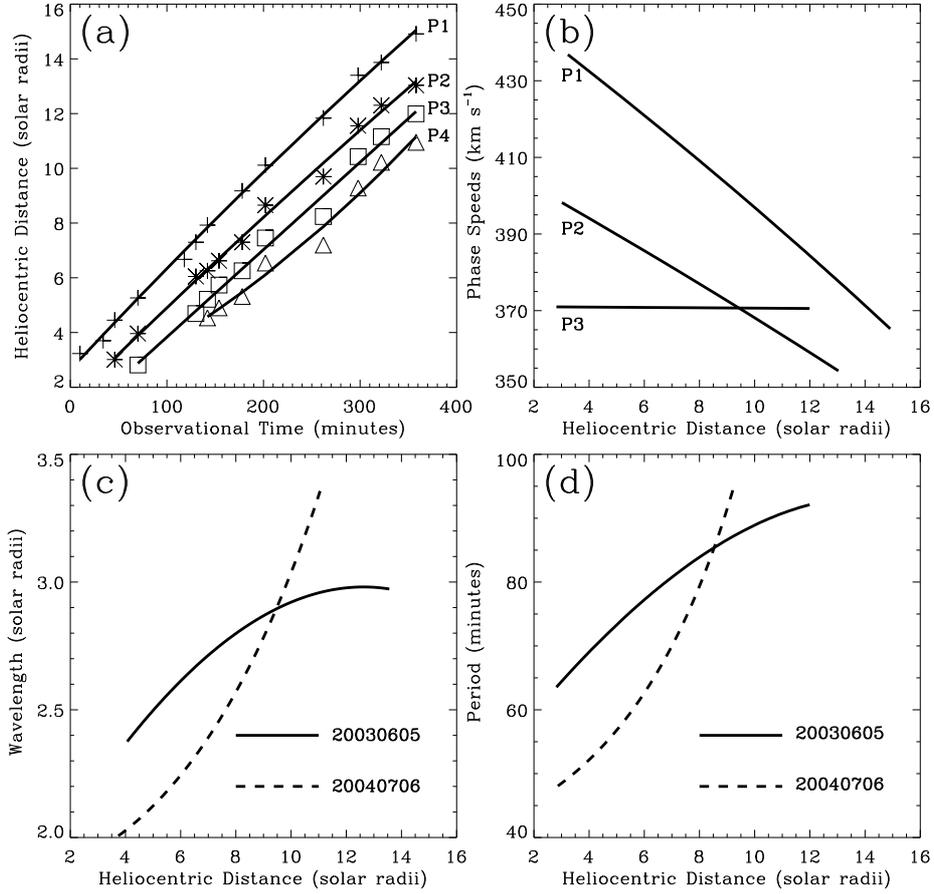}
              }
              \caption{Properties of the streamer wave event
              20030605. (a) Radial variations of heliocentric
              distances of the four phases (P1 - P4) indicated by plus signs,
              asterisks, squares, and triangles, respectively. The solid lines
are given by the second order polynomial fittings to the
measurements. (b) The fitted profiles of phase speeds for the
first three wave phases. The deduced radial evolutions of the
wavelengths (c) and periods (d) for the 20030605 (solid) and the
20040706 (dashed) events.}
   \label{f8}
   \end{figure}
\begin{figure}    
   \centerline{\includegraphics[width=1\textwidth]{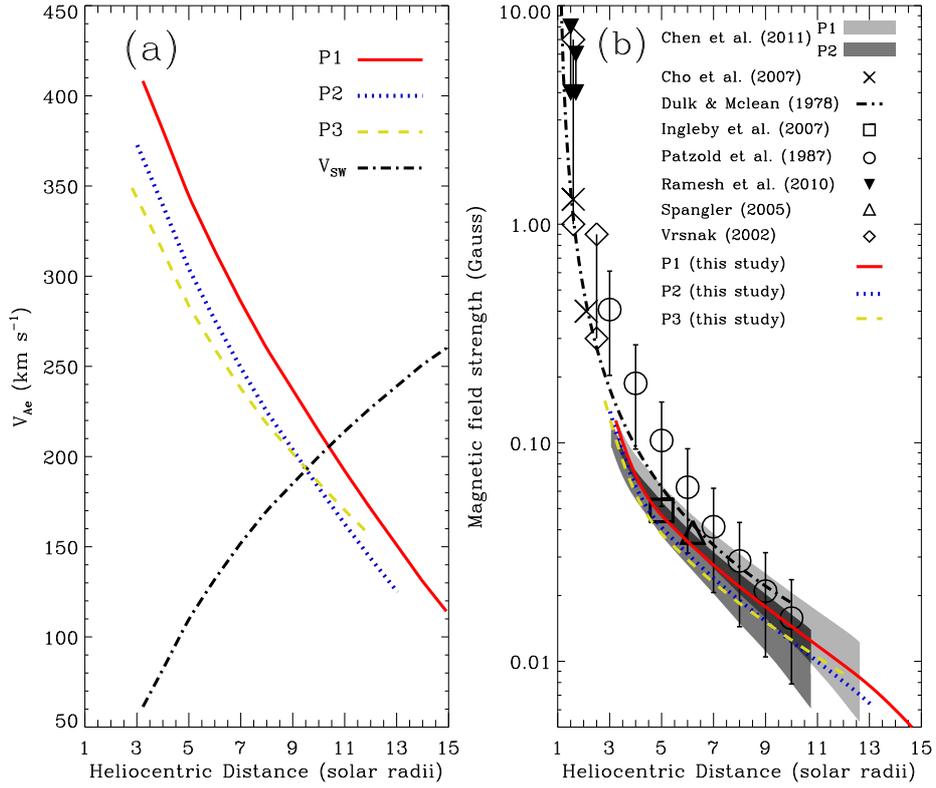}
              }
              \caption{Radial profiles of the estimated Alfv\'en speed (a) and magnetic
              field strength (b) in the region surrounding
              the streamer plasma sheet for the streamer wave event 20030605. Profiles corresponding
              to different wave phases are given by the red-solid (P1), blue-dotted (P2), and
              yellow-dashed (P3) lines. The black dot-dashed line in (a) presents the average blob speed
              taken from Wang et al. (2000). Various symbols in (b) represent other estimates of the
              coronal magnetic field strength.
                      }
   \label{f9}
   \end{figure}
\begin{figure}    
   \centerline{\includegraphics[width=1\textwidth]{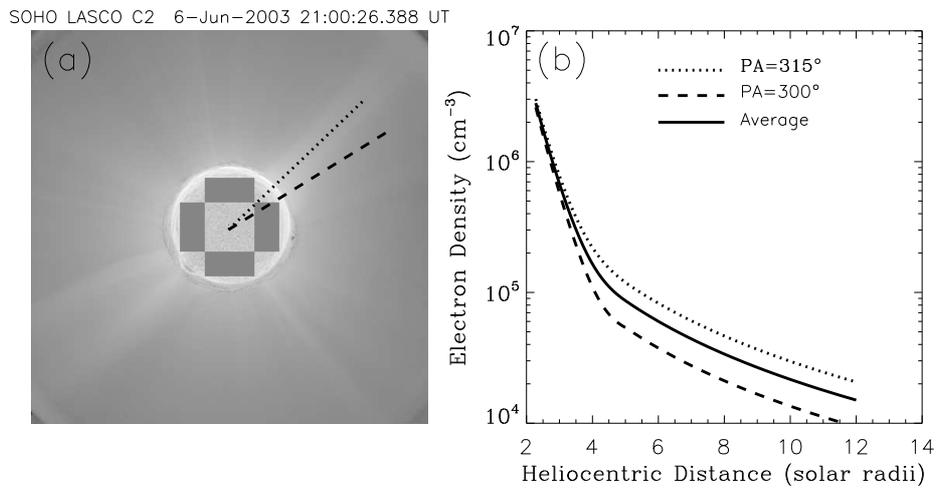}
              }
              \caption{(a) The pB intensity distribution observed
              by LASCO on 6 June 2003, the dotted (dashed) line
              denotes a specific position angle of 315 (300)$^{\circ}$. (b)
              Dotted and dashed lines are the corresponding electron number
              density profiles given by the pB inversion method ($\le$ 5 R$_\odot$) and
              the r$^{-2}$ dependence ($>$ 5 R$_\odot$) along the
              two position angles. The solid line is given by their average.}
   \label{f10}
   \end{figure}
\end{article}
\end{document}